\documentstyle[proceedings,epsf]{crckapb}

\def\gtorder
{\mathrel{\raise.3ex\hbox{$>$}\mkern-14mu\lower0.6ex\hbox{$\sim$}}}
\def\ltorder
{\mathrel{\raise.3ex\hbox{$<$}\mkern-14mu\lower0.6ex\hbox{$\sim$}}}

\def\mnras	{{\it M.N.R.A.S.\/}}
\def\apjs	{{\it Ap.J.\/Supp.\/}}
\def\apj	{{\it Ap.J.\/}}
\def\aj		{{\it A.J.\/}}
\def\aap	{{\it A. \& A.\/}}

\def 		\etal   {{et al.\thinspace}}

\def 		\cf     {{cf.}}

\def 		\kms    {{\rm km s$^{\hbox{\tiny --1}}$}}

\def\lsim	{ \rlap{\lower .5ex \hbox{$\sim$} }{\raise .4ex \hbox{$<$} } }
\def\gsim	{ \rlap{\lower .5ex \hbox{$\sim$} }{\raise .4ex \hbox{$>$} } }

\def\msolar	{ \rm {M_{\odot}} }
\def\HI		{{H{\sc I}}}
\def\etal	{{\it et~al.}}

\begin{opening}
\title{The Distribution of Dark Mass in Galaxies}
\subtitle{Techniques, Puzzles, and Implications for Lensing}

\author{Penny D. Sackett}
\institute{Kapteyn Astronomical Institute\\
           9700 AV Groningen, The Netherlands}

\end{opening}

\runningtitle{DISTRIBUTION OF DARK MASS IN GALAXIES}

\begin{document}



\section{Abstract}

Gravitational lensing is one of a number of methods
used to probe the distribution of dark mass in the Universe.
On galactic scales, complementary techniques include the use of
stellar kinematics, kinematics and morphology of the neutral gas layer,
kinematics of satellites, and morphology and temperature profile
of X-ray halos.  These methods are compared, with emphasis on their
relative strengths and weaknesses in constraining
the distribution and extent of dark matter in the Milky Way and
other galaxies.  It is concluded that (1) the extent of dark halos
remains ill-constrained, (2) halos need not be isothermal,
and (3) the dark mass is probably quite flattened.

\section{Introduction}

Modeling the gravitational structure of a galaxy, and therefore its
lensing properties, requires knowledge of the extent,
radial profile, and geometric form of its mass distribution.
The interpretation of microlensing rates and optical
depths along different lines of sight through the Milky Way,
for example, is strongly dependent on the assumed distribution of
total (light and dark) Galactic mass.  On larger
scales, efficient and reliable image-inversion techniques designed
to measure the structure parameters of intervening lensing galaxies
require appropriate fitting functions for the lensing mass.

This review focuses on techniques that
form a symbiotic relationship with lensing in producing
valuable and complementary constraints on galactic potentials,
especially in providing partial answers to the
following questions about galactic {\it dark\/} mass:
\begin{itemize}
\item What is the physical extent of dark matter in galaxies?
\item Is the distribution of dark mass isothermal?
\item What is the shape of dark ``halos''?
\end{itemize}

\section{How Big are Dark Halos?}

The size of dark halos controls the galactic ``sphere of influence''
for lensing, interactions, and accretion.
Together with the radial and vertical structure parameters,
halo extent determines the total mass of the galaxy.
The notion of halos as distinct entities ceases
to be useful, of course, on scales larger than half the mean
distance to the nearest, comparably-sized neighbor.

If halos are extremely large
and isothermal, they cannot be totally baryonic without violating
the constraints on $\Omega_B h^2$
from primordial big bang nucleosynthesis (BBN) models.  Recent
assessments give $0.01 \leq \Omega_B \leq 0.06$ for $0.5 \leq h_{100} \leq 1$
(Walker \etal\ 1991, Smith, Kawano \& Malaney 1993).
Since the density of observed baryons is $\Omega_{\rm Lum} \ltorder 0.007$
(Pagel 1990), the average ratio of dark-to-luminous baryons is
$0.4 \leq M_{B,\rm Dark}/M_{B,\rm Lum} \leq 8$,
and at least some of the Universe's baryons are dark.
Rotation curve analysis indicates that
$1 \ltorder M_{B,\rm Dark}/M_{B,\rm Lum} \ltorder 10$ (\cf\ Broeils 1992),
so that on scales comparable to \HI\ disks ($\sim$30~kpc),
halos composed entirely
of dark baryons are consistent with BBN.
Faint galaxies are more numerous and more dark matter dominated
than brighter galaxies, but contribute less to the
total luminosity of the Universe.
Thus the upper limit placed by BBN on the size of
baryonic halos is likely to be considerably larger than 30~kpc
(Binney \& Tremaine 1987), but its
calculation requires a model-dependent integral over the
galaxy luminosity function, weighted by $M_{B,\rm Dark}/M_{B,\rm Lum}$.

\subsection{The Extent and Mass of the Milky Way Halo}

At large radius, the mass of our galaxy can be estimated
from the kinematics of distant, presumably bound, objects --- halos
stars, satellite galaxies, and group members ---
and from the kinematics of
Magellanic Clouds/Stream system. The former has been done most
recently by Kochanek (1995), who finds that, using a Jaffe model
as the global mass distribution for the Galaxy,
the total mass inside 50~kpc at 90\% confidence
is $(5.4 \pm 1.3) \times 10^{11} \msolar$
if the timing constraints of the Local Group are imposed
and Leo I is bound, and somewhat lower at
$4.3^{+1.8}_{-1.0} \times 10^{11} \msolar$
if the timing constraints are not imposed.
The corresponding masses within 100~kpc are
$7^{+4}_{-3} \times 10^{11} \msolar$ and
$(8 \pm 2) \times 10^{11} \msolar$, respectively.
A recent re-examination of the kinematics and proper motions of
the Magellanic Clouds and Stream using two different
model potentials for the Milky Way (Lin, Jones \& Kremola 1995)
yields $(5.5 \pm 1) \times 10^{11} \msolar$ inside
100~kpc.  About one-half this mass must lie {\it outside\/} the
present Cloud distance (50~kpc) in order to explain with
these models the observed infall of the Magellanic Stream.

Since the luminous matter in the Galaxy accounts
for $(0.6-1) \times 10^{11} \msolar$,
the full range of dark mass estimates from these two methods is
$1.3 \times 10^{11} \ltorder M_{\rm Dark}(<50 \rm kpc) \ltorder
6.1 \times 10^{11} \msolar$,
with apparent contradictions at the lower end with
Local Group timing and at the upper end with Magellanic
Stream kinematics.
The implied upper limit of $M_{\rm Dark}(<50 \, \rm kpc) \sim2.7
\times 10^{11} \msolar$ from the Magellanic Stream model is
only just consistent with the lower limit of $\sim2.3 \times 10^{11} \msolar$
from the satellite model.
For comparison, the spherical isothermal
dark halo used by many microlensing teams as a fiducial model
contains $4.1 \times 10^{11} \msolar$
interior to the LMC distance of 50 kpc (Griest 1991).
Using this model and its first year's LMC data, the MACHO team
concludes with 68\% confidence that the
total mass in compact dark lenses is $7^{+6}_{-4} \times 10^{10} \msolar$
(Alcock \etal\ 1995).
(For more complete reviews of the mass of the Galaxy and its dependence
on assumptions about Leo I, see Fich \& Tremaine 1991, Schechter 1993, and
Freeman, these proceedings.)

\subsection{Halo Size of External Galaxies}

The kinematics of satellites can also be used to study
the halos of external galaxies, but since only a small number of
satellites are observed per primary,
conclusions are based on a statistical analysis of the
sample as a whole.  Based on satellite velocities and \HI\ rotation curves,
Erickson, Gottesman and Hunter (1987) concluded that
the primaries in their sample have $M_{Dark}/M_{Lum} < 5$,
total $M/L \sim 20$, and potentials that
are well-described by a point mass model --- all consistent
with dark halos that extend no more than 3 disk radii.
In a more recent study using a different sample, however,
Zaritsky and White (1994) conclude that halos are nearly isothermal, with
total $M(<200 \rm kpc) = 1.5-2.6 \times 10^{12} \msolar$
and $110 < M/L < 340$ (for $h_{100}=3/4$).
Their result is primarily due to secondaries at 200-300~kpc,
where the orbital times are on the order of a Hubble time, thus
necessitating the use of halo formation models
to interpret the satellite kinematics.
Using their method, Zaritsky and White conclude that
the Erickson \etal\ sample,
which has smaller mean primary-satellite separation,
is consistent with both small and large mass halos.

In the future, weak lensing is likely to play a larger
role in constraining the extent of dark halos.  Recent work by
Brainerd, Blandford and Smail (these proceedings)
has given the first indication that the tangential distortion of background
galaxies due to weak lensing by foreground galaxies is statistically
measurable for a large sample ($\sim 3000$) of source-lens pairs.
Their measurement
of $1.0^{+1.1}_{-0.7} \times 10^{12} h^{-1} \msolar$
for the total mass within 100 $h^{-1}$ kpc is consistent both
with a mass distribution that grows linearly to 100~kpc
and one that truncates much sooner with total $M/L \approx 10$.

The ring of \HI\ gas in the M96 group in Leo (Schneider 1985) offers
the rare opportunity to sample galactic potentials at very large radii
using the well-defined orbits of cold gas.  The Leo ring has a
radius of 100 kpc and completely encircles the early-type
galaxies M105 and NGC~3384.
The radial velocities and spatial distribution
of the gas are consistent with a single, elliptical {\it Keplerian\/}
orbit with a center-of-mass velocity equal to the centroid of
the galaxy pair, and a focus that can be placed at the barycenter of
the system without compromising the fit.  The implied
dynamical mass within 100~kpc is $5.6 \times 10^{11} \msolar$,
(only twice that inferred from the internal dynamics of the galaxies),
giving a total $M/L \approx 25$.
The sensitivity of non-circular orbits
to the power law form of the potential suggests that dark
matter does not extend much beyond the ring pericenter
radius of 60~kpc.
As a caveat, it is yet clear to what degree M96, a spiral
located 60~kpc (in projection) outside the ring, may perturb
the ring kinematics.

\section{Are Dark Halos Isothermal?}

The approximate flatness of \HI\ rotation curves
is the best observational evidence that dark matter is
present in spirals and has a shallower density profile than
the light; an isothermal halo is
as shallow as $r^{-2}$.  Early theoretical studies (Gott 1975)
suggested that violent relaxation would cause the inner regions of
galaxies to have steeply falling profiles that
would flatten to $r^{-2.25}$ in the outer parts.
More recent CDM models (Navarro, Frenk \& White 1995)
indicate that dark halo profiles may be shallower than $r^{-2}$ in
the center and quite steep near the virial radius.
Compression by a dissipating gaseous disk
may further contract and flatten
the dark matter (Blumenthal \etal\ 1986), accounting in part for
the apparent ``conspiracy'' between the
dark and luminous mass that produces flat rotation curves.

\subsection{Milky Way}

The radial structure of the mass and light in
the Milky Way is less well-constrained than in external
galaxies.  Determining the rotation curve of the Galaxy, in particular,
has proven notoriously difficult.
On the other hand, distances and kinematics of old, resolved stars
can be used to measure the vertical restoring force
of the local disk --- and thus its surface mass density.
In this way, Kuijken and Gilmore (1991) report a mass column of
$71 \pm 6 \msolar pc^{-2}$ within a 1.1~kpc band from the
Galactic plane,
with $48 \pm 9 \msolar pc^{-2}$ due to the disk itself, and
the rest contributed by a rounder halo.  Other recent estimates
are similar: Gould (1990) weighs in at $54 \pm 8 \msolar pc^{-2}$,
Bahcall, Flynn and Gould (1992) at $54 \pm 8 \msolar pc^{-2}$, and
Flynn and Fuchs (1994) at $52 \pm 13 \msolar pc^{-2}$.
The dynamical disk mass thus seems to be in
remarkable agreement with the detectable disk mass of
$49 \pm 9 \msolar pc^{-2}$ --- at least locally,
almost none of the disk mass is dark.
Since only about one-half of the local rotation support
is provided by the observable disk, this further implies that
dark matter in the Galaxy is dynamically important
at radii as small as 2.5 disk scale lengths.
Stated in the language of \S4.2,
the Milky Way disk is one-half of its ``maximal disk'' value.
Unfortunately, uncertainties in the outer Galactic rotation
curve frustrate attempts to determine the distribution of mass in
the outer Galaxy, which is further complicated by a recent suggestion that the
generally-accepted local rotation speed, $\Theta_0 = 220$ \kms,
may be overestimated by $\sim$10\% (Merrifield 1992).
A smaller value would increase the relative dynamical importance
of the luminous disk and decrease the slope of the outer
rotation curve, to which $\Theta_0$ is tied.

Conclusions drawn from microlensing results about the
dark baryonic content of the Milky Way
depend on the assumed distribution
of dark {\it and\/} luminous matter in the Galaxy
(\cf\ Paczy\'nski, these proceedings).
Many studies have explored how different assumptions for $M/L$,
rotation curve slope, and the shape, truncation radius and
radial profile of the halo affect these
conclusions (\cf\ references in Griest \etal\ 1995).
As an indication of the importance of {\it luminous\/} structure,
lensing by stellar bars in the Milky Way and the LMC
has been held accountable, respectively, for most of the optical depth toward
the Galactic center (Zhao, Spergel \& Rich 1995) and the
LMC (Sahu 1994).  On the other hand, if the Galactic disk were ``maximal, ''
the MACHO results toward the LMC would be consistent with a dark halo
entirely composed of lensing baryons
(Alcock \etal\ 1995).

\subsection{Radial Distribution of Dark Mass in External Galaxies}

In contrast to the difficulties in the Milky Way,
surface brightness profiles and rotation curves for external galaxies
can be measured well,
but their disk mass-to-light ratios, $M/L$, are uncertain.
A disk $M/L$ that is constant with radius (but varies from galaxy to
galaxy) can explain the kinematics within the optical radius of
many spirals (\cf\ Kalnajs 1983, Kent 1986, Buchhorn 1992), but
the high velocities observed at the edges of \HI\ disks can be
reproduced only by invoking a rapid radial increase in $M/L$
(\cf\ Kent 1987, Begeman 1987).
Since the age and metallicity gradients inferred from the
blueing radial color gradients in spirals do not
produce these strong, {\it positive\/} gradients in $M/L$
(\cf\ de~Jong 1995), dark matter is implicated.

In order to estimate conservatively the amount of
dark matter in a galaxy, the ``maximum disk hypothesis'' is often adopted
(van Albada \& Sancisi 1987), which fixes the disk $M/L$
at the value that maximizes the disk mass
without violating kinematic constraints.
The hypothesis is controversial (\cf\ Rubin 1987,
Casertano \& van Albada 1990, Freeman 1993),
but when it is used to fit rotation curves, the resulting disk $M/L$
are larger for brighter and earlier type spirals than for
fainter and later type spirals (Broeils 1992, Buchhorn 1992).
The correlation appears to be stronger in bluer bands.
These trends may be due to the older stellar populations
associated with early spirals,
a notion supported by comparison with the $M/L$ derived
from stellar population synthesis models
(Athanassoula, Bosma \& Papaioannou 1987) and the observed
stellar dispersions in spirals (van der Kruit \& Freeman 1986).
Alternatively, they may reflect trends in dark matter properties
with galaxy type and luminosity that are incorrectly characterized by
the application of maximum disk models (van der Kruit 1995).

\begin{figure}
\vskip -1.3cm
\epsfxsize=\hsize\epsffile{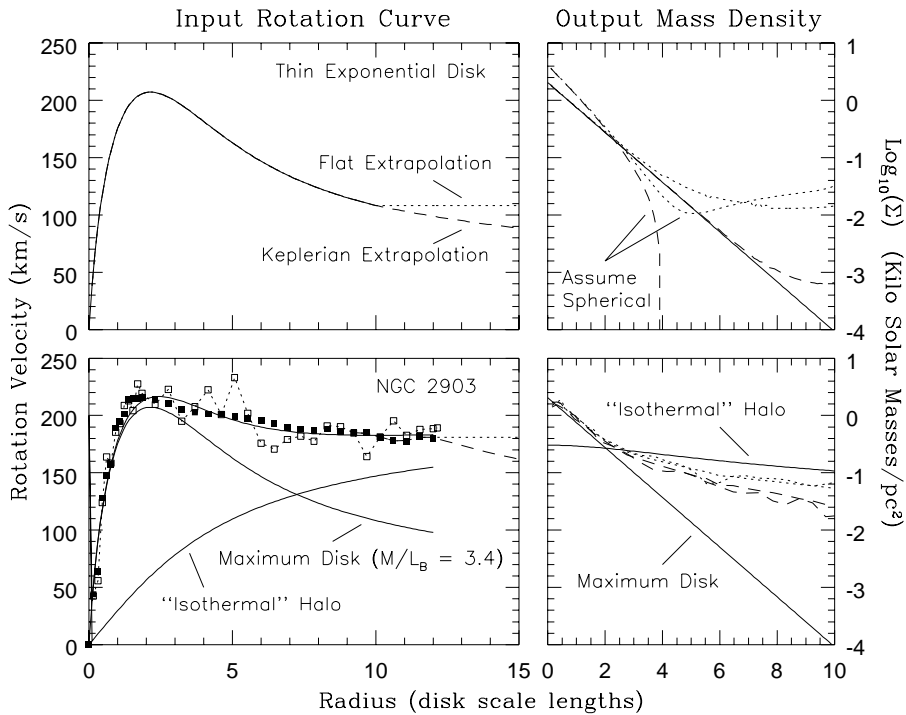}
\vskip -2.0cm
\caption{
Inversion of rotation curves (left)
to derive dynamical surface mass densities, $\Sigma$ (right).
Two extrapolations for $v(r)$ are shown: flat
(dotted) and Keplerian (dashed).
{\it Top: Thin, exponential disk.\/} Both extrapolations overestimate
the true $\Sigma$ (solid) because for an
exponential disk $v(r)$ declines faster than Keplerian.
Only for inner half of the disk can $\Sigma$ be determined reliably.
A spherically-symmetric mass estimator is unreliable for an exponential
disk and can produce negative $\Sigma$ in the inversion.
{\it Bottom: Sbc NGC~2903.\/} Real data (solid squares) and
artificially noisy data (open squares) are
inverted using both extrapolation schemes.
$M/L \, $ increases markedly beyond $\sim$2 scale lengths.
Adding noise makes little difference.   A spherical isothermal
halo has a $\Sigma$ that is $\sim\pi/2$ larger than that
of the flat extrapolation, but with similar slope (Sackett, in preparation).
}
\vskip -0.5cm
\end{figure}

It should be stressed that {\it rotation curves do not
constrain the dark matter
distribution to have a $r^{-2}$ (isothermal) volume density profile\/}.
If (1) rotation curves were perfectly flat, (2) halos were spherical,
and (3) the luminous mass were negligible,
then indeed dark matter halos could be described by
singular isothermal spheres over the radial range of the kinematics.
In fact, rotation curves are seldom flat,
but instead have slopes that are
systematically related to the peak speed or the luminosity
concentration of the galaxy  (Kent 1987, Athanassoula, Bosma \&
Papaioannou 1987, Casertano \& van Gorkom 1991, Broeils 1992):
diffuse, slow rotators have rising rotation curves,
while compact, fast rotators have falling curves.
Evidence is mounting that dark halos are not spherical (\S 5).
Finally, since the stellar mass is not strongly constrained,
dark halos with asymptotic $r^{-3}$ and $r^{-4}$ density profiles
are also consistent with observed rotation curves (Lake \& Feinswog 1989).
Even when $r^{-2}$ halos are used to fit rotation curves ---
together with luminous disks of reasonable $M/L$ ---
a large core radius must be assumed, so that the halo does not
achieve its asymptotic (isothermal) speed at the last measured point
(Fig.~1).  This is especially true of maximum disk fits
(\cf\ Broeils 1992), but often applies to fits that assume
smaller disk masses as well (\cf\ Kent 1987).
This suggests that the linewidth of the \HI\ gas used
in the Tully-Fisher relation is probably not governed
by the asymptotic speed of an isothermal dark halo.

Rotation curve inversion is a step toward a model-independent
method for determining the radial distribution of dark mass in galaxies.
The technique has been criticized as being sensitive to noise (Binney \&
Tremaine 1987), but for this application the typical uncertainties of 10-20\%
are quite tolerable.
The method does depend on the assumed geometry
of the mass and extrapolation of the rotation curve beyond the
last measured point (Fig.~1), but has the advantage of making
this dependence explicit rather than camouflaging it by the use of
a particular model for the dark mass.

\section{What is the Shape of the Dark Mass: Disks or Halos?}

Use of term ``halo'' to describe
the distribution of dark matter may be prejudicial: there is
no strong theoretical or observational evidence to indicate that
dark matter in galaxies is distributed spherically.
Dark halos have been favored over dark disks as a means
to stabilize galaxies against bar formation (Ostriker \& Peebles 1973),
but bulges (Kalnajs 1987) and hot disks (Athanassoula \& Sellwood 1987)
are now believed to be more efficient stabilizers.
Traditional rotation curve analysis is insensitive to vertical
structure, but accumulating observational evidence from other
methods suggests that the dark mass may be considerably flattened toward the
stellar plane, while remaining relatively axisymmetric,
with an in-plane axis ratio of $(b/a)_\rho > 0.7$
(see review by Rix 1995).

Here, we focus on $(c/a)_\rho$, the vertical flattening of
dark matter, since it is likely to have the stronger implications for
both microlensing in the Galaxy
(\cf\ Gould, Miralda-Escud\'e, \& Bahcall 1994),
and the use of macrolensing as a probe of galaxy structure.
The flattening of the dark mass may also provide a clue
as to the nature of its constituents.
N-body simulations of dissipationless collapse produce strongly
triaxial dark halos (Frenk \etal\ 1988,
Dubinski \& Carlberg 1991, Warren \etal\ 1992), but adding a
small fraction ($\sim$10\%) of dissipative gas results in halos
of a more consistent shape --- nearly oblate, $(b/a)_\rho \gtorder 0.8$,
but moderately flattened, $(c/a)_\rho \gtorder 0.6$
(Katz \& Gunn 1991, Dubinski 1994).
Thus strongly flattened halos, with $(c/a)_\rho < 0.5$, may imply that
dissipation has played an even greater role, perhaps implicating
baryonic dark matter.

In order to measure $(c/a)_\rho$ of
the dark mass, a probe of the vertical gradient of the potential
is required.
In the Milky Way, the measured anisotropy of the velocity dispersion
of extreme Population II halo stars has been used to estimate
the flattening of the mass distribution (Binney, May \& Ostriker 1987,
van~der~Marel 1991).  Unfortunately, the results depend on the
unknown orbital structure of the stellar halo, so that $(c/a)_\rho$ can
be confined only to lie between 0.3 and 1 at the solar
neighborhood.

In external galaxies, Buote and Canizares (1994, 1995) have used
the flattening of extended X-ray isophotes, assuming
that the gas is in hydrostatic equilibrium, to place constraints on
the flattening of the dark matter in two early-type systems.
For the elliptical NGC~720, they find that the dark isodensity contours
have axis ratio $0.3 \ltorder (c/a)_\rho \ltorder 0.5$ at 90\% confidence;
for the lenticular NGC~1332, $0.2 \ltorder (c/a)_\rho \ltorder 0.7$.
This suggests that these dark halos are at least as flattened as
their corresponding luminous galaxies, which have optical isophotes
of axis ratio $q \approx 0.6$.

These values contrast with that of $(c/a)_\rho \geq 0.84$
derived for the S0 NGC~4753
by Steiman-Cameron, Kormendy \& Durisen (1992)
on the basis of fitting an inclined,
precessing disk model to the complicated pattern of the
galaxy's dust lanes.  Their remarkably good fit
is independent of $(c/a)_\rho$; the flattening
constraints are based on the assumption that
the gas is smoothly distributed and has completed at least 6
orbits at all radii.

Stable rings around galaxies are not observed to have random
orientations, but are found preferentially close to
the equatorial or polar planes, suggesting that the potential
may be oblate.
In particular, polar ring galaxies (PRGs) are surrounded
by rings of gas and stars in orbits nearly perpendicular to the
central stellar plane; these rings can extend to
20 disk scale lengths.
Since in an oblate potential closed ring orbits are elongated
along the polar axis and have speeds that vary with ring azimuth,
the shape and kinematics of a polar ring are excellent
extended probes of $(c/a)_\rho$.
Early kinematic analyses of three PRGs
produced axes ratios for the {\it potentials\/} of
$0.86 < (c/a)_\Phi < 1.05$ with uncertainties of 0.2
(Schweizer, Whitmore \& Rubin 1983, Whitmore, McElroy \& Schweizer 1987),
corresponding to $0.58 \ltorder (c/a)_\rho \ltorder 1.15$
with very large uncertainties.
Subsequent studies using more detailed
mass models and higher quality data over a larger radial
range have narrowed the range for the dark mass to
$0.3 \ltorder (c/a)_\rho \ltorder 0.6$ (Arnaboldi \etal\ 1993, Sackett
\etal\ 1994, Sackett \& Pogge 1995);
in each galaxy, $(c/a)_\rho$ is
similar to the inferred flattening of the central {\it stellar\/} body.

Measurements of $(c/a)_\rho$ for spiral galaxies are rarer,
more difficult, and sorely needed.
Assuming that gas disks evolve gravitationally toward a
discrete bending mode in tilted rigid halos,
Hofner and Sparke (1994) find that moderate halo flattening
of $0.6 \ltorder (c/a)_\rho \ltorder 0.9$ can
reproduce the observed \HI\ warps of five spirals.
In principle, $(c/a)_\rho$~can also be constrained by
the flaring of the \HI\ layer; in the most detailed study of this type,
Olling and van~Gorkom (1995) obtain $0.2 < (c/a)_\rho < 0.8$
for the dark halo of the Sc NGC~4244.
Since non-gravitational energy sources may be responsible for
a substantial fraction of the vertical support of gas
(Malhotra 1995, and references therein), this measurement
may be an upper limit to $(c/a)_{\rho}$.

\section{Parting Caveats and a Puzzle}

Since the mass distribution in cluster galaxies
may be modified by the interactions and
violent relaxation that shape the evolving cluster potential,
we have restricted this review to relatively
isolated galaxies that are more likely to be dynamically relaxed.
Furthermore, we have largely ignored ellipticals,
the inner few kpc of which are thought to be responsible
for the strong lensing of distant QSOs and radio sources.
Although selection effects operate to favor
flattened lenses in multiply-imaged systems (Kassiola \& Kovner 1993),
image inversion techniques yield lenses
that are surprisingly flattened (Kochanek 1995a, and
references therein) --- the {\it projected\/} $(c/a)_{\Phi} \ltorder 0.8$
corresponds to $(c/a)_\rho \ltorder 0.4$.
Can these flat lenses be reconciled with the axis ratio distribution of
ellipticals, which peaks at $q = 0.7$ (Ryden 1992), or are disk galaxies
implicated?

\vskip 0.3cm
\noindent\small{
P.D.S. gratefully acknowledges travel support from
the Leids Kerkhoven-Bosscha Fonds and the International Astronomical Union.
}

\end{document}